\documentclass[seceq]{ptptex}

\usepackage{graphicx}

\title{
Five Point Mass Gravitational Lenses in a Rhombus 
as a Soluble Model Giving the Maximum Number of Images %
}

\author{
Hideki \textsc{Asada}%
}

\inst{
Faculty of Science and Technology, Hirosaki University, 
Hirosaki 036-8561, Japan
}

\abst{
As an extension of four point mass lenses at the vertices 
of a rhombus, we present five point mass lenses 
at the center and vertices of a diamond, 
which constitute, for a source behind the center, 
a soluble model giving expressions of all the image positions 
(with the maximum number of images as twenty for five lenses). 
For a source near the center, 
all the image positions are obtained in the linear approximation. 
}

\begin{document}

\maketitle

\section{Introduction}
Since the discovery of the lensed quasar Q0957+561 in 1979 
(Ref. \citen{WCW}, 
the gravitational lensing has been used to find 
distant objects and determine cosmological parameters 
such as the cosmological constant \cite{FFKT} and 
inhomogeneities (e.g., Ref. \citen{YKN} 
and 
astrophysical ones 
such as the mass of MACHOs (e.g., Ref. \citen{MACHO}. 

In spite of its importance, there are unsolved problems 
in the gravitational lens theory 
because of its nonlinear nature. 
One is that roots for the so-called lens equation 
(which correspond to image positions) 
are unknown in analytic forms, 
and numerical roots can be obtained, 
except for very few cases. 
Very recently, perturbative methods have been 
provided for obtaining the roots that can be expressed 
as Maclaurin series in mass ratios \cite{Asada09}. 
Positions of extra images that cannot be expressed 
in Maclaurin series are still unknown 
as functions of lens parameters. 

Even two point masses are too complicated 
to treat by hand, since the lens equation 
is in fifth order in real variables \cite{Asada02}
as well as complex ones \cite{Witt90}. 
Fifth order in complex variables could 
lead to higher orders in real ones but 
this is not the case. 
Only for the special position of the source, 
the binary lens equation becomes exactly solvable \cite{ES,SW}. 
Until recently, it has been unknown how many 
the maximum number of images for $N$ point mass lenses is. 
Eventually\cite{Rhie01,Rhie03,KN06}, Rhie showed that it is $5(N-1)$. 
The maxima are achieved indeed by a binary as five. 
One important example is four lenses at the vertices 
of a rhombus but with numerical methods \cite{Rhie01}. 
This example was considered as an extension of 
the earlier work by Mao, Petters and Witt (Ref. \citen{MPW}, 
where they investigated properties of 
point mass lenses on a regular polygon with $N\geq3$, 
though for this model the number of images 
is not more than $3N+1$ (less than $5(N-1)$) and 
thus does not provide the maximum number. 
Rhie (2003) first considered a model of N equal masses 
on a regular polygon and next added a small mass at the center 
in order to prove that the maximum number of images is no less than
$5(N-1)$ without giving expressions for the image positions 
\cite{Rhie03}. 

The purpose of this brief paper is to present 
certain soluble models for gravitational lenses. 
First, we shall reexamine, especially in exact algebraic manners, 
four point mass gravitational lenses 
at the vertices of a rhombus. 
Next, as an extension, we investigate 
five point mass lenses 
at the center and vertices of a diamond, 
which constitute, for a source located behind the center, 
a soluble model giving expressions of all the image positions 
(with the maximum number of images as twenty for $N=5$). 
In this sense, the present paper can be considered 
as an extension of the previous papers 
\cite{Rhie01,Rhie03}.

\section{Soluble models giving the maximum number of images}
The lens equation, expressed in complex quantities \cite{BKN,BK}, 
is given for $N$ point mass lenses by \cite{Witt90} 
\begin{equation}
w =z - \sum_{p=1}^N \frac{m_p}{z^* - \epsilon_p^*} , 
\label{lenseq-z}
\end{equation}
where the source, image and lens positions   
are denoted as $w=\beta_x+i \beta_y$, 
$z = \theta_x + i \theta_y$, 
$\epsilon_p = e_{p x} + i e_{p y}$, respectively, 
the mass ratio is denoted as $m_p$, 
and the asterisk $*$ means the complex conjugate. 
Here, the angle is normalized by the Einstein ring radius 
defined by the total mass.

\subsection{Four lenses at the vertices of a rhombus}
Let equal masses ($m_i=1/4$) located at the vertices of a rhombus 
(See the next subsection for inequal mass cases), 
where its center is chosen as the center of the coordinates 
and the location of each mass (in the anti-clockwise direction) 
is denoted as 
$\epsilon_1=k$, $\epsilon_2=i \ell$, 
$\epsilon_3=-k$, $\epsilon_4=-i \ell$ 
for $k, \ell \in R$. 
Without loss of generality, we assume $k\geq\ell$. 

Let us put the source at the center 
(i.e. the maximally symmetric location). 
Then, the equation is rewritten as 
\begin{equation}
z (z^{* 2}-k^2) (z^{* 2}+\ell^2) 
= z^* 
\left( z^{* 2} + \frac12 (\ell^2-k^2) \right) .  
\label{lenseq-four}
\end{equation}
We employ the polar coordinates as $z= r \exp(i\phi)$. 

An immediate solution for Eq. ($\ref{lenseq-four}$) 
is $r=0$. 
We shall seek other solutions with $r \neq 0$. 
Eq. ($\ref{lenseq-four}$) is reexpressed as 
\begin{equation}
r^4 - r^2 = A_4 e^{4i\phi} + B_4 e^{2i\phi} , 
\label{lenseq-four-r}
\end{equation}
where real quantities are defined as 
\begin{eqnarray}
A_4&=&k^2 \ell^2 , 
\label{A4}
\\
B_4&=&(\ell^2 - k^2) \left(\frac12 - r^2\right) . 
\label{B4}
\end{eqnarray}
Here, the L. H. S. of Eq. ($\ref{lenseq-four-r}$) is real and 
hence the R. H. S. must be real. 
We thus have three cases as 
$\exp(2i\phi)=\pm 1$ 
or $A_4[\exp(2i\phi)+\exp(-2i\phi)]+B_4=0$. 

\noindent
\underline{Case 1}. 
$\exp(2i\phi)= 1$ 
($\phi = 0$ or $\pi$):\\  
The lens equation becomes 
\begin{equation}
r^4 + (\ell^2 - k^2 -1) r^2 + \frac12 (k^2 - \ell^2) - k^2 \ell^2 = 0 , 
\label{lenseq-four-r1}
\end{equation}
where the coefficient of $r^2$ is always negative 
because of $k \geq \ell$. 
This equation quadratic in $r^2$ can be solved immediately. 
According to Descartes' rule of signs (e.g., Ref. \citen{Waerden}, 
which states that the number of positive roots either equals to 
that of sign changes in coefficients of a polynomial or 
less than it by a multiple of two, 
we have the maximum number of positive roots 
as $2$ 
if $(k^2 - \ell^2)/2 > k^2 \ell^2$, 
whereas we have only one positive root otherwise.
It is worthwhile to mention that the number of the positive roots 
can be precisely counted by using an explicit form of the roots, 
though it needs straightforward but lengthy calculations 
(especially for a cubic equation discussed later).

\noindent
\underline{Case 2}.  
$\exp(2i\phi)= -1$ 
($\phi = \pi/2$ or $3\pi/2$):\\  
The lens equation becomes 
\begin{equation}
r^4 + (k^2 - \ell^2 -1) r^2 + \frac12 (\ell^2 - k^2) - k^2 \ell^2 = 0 , 
\label{lenseq-four-r2}
\end{equation}
where the sum of the constant terms is always negative 
for $k \geq \ell$. 
This equation for $r^2$ can be solved immediately. 
Regardless of the sign of the coefficient of $r^2$, 
the number of positive roots is one 
according to Descartes' rule of signs. 

\noindent
\underline{Case 3}.  
$A_4[\exp(2i\phi)+\exp(-2i\phi)]+B_4=0$:\\ 
Given $r$, this provides four roots of $\phi$ (mod $2\pi$). 
For this case, Eq. $(\ref{lenseq-four-r})$ is reduced to 
\begin{equation}
r^4 - r^2 + k^2 \ell^2 = 0 . 
\label{lenseq-four-r3}
\end{equation}
If and only if $k\ell < 1/2$, 
we have two positive roots. 

Therefore, we have the maximum number of images as 
$1 + 2\times 2 + 2\times 1 + 4\times 2 = 15$ 
if the above conditions are satisfied. 
The number is exactly the maximum number of images for $N=4$. 

A rhombus with right angles is a square, 
for which there is a change in the Case 1. 
The sum of the constant terms becomes negative. 
According to Descartes' rule of signs, 
the number of positive roots is necessarily one. 
As a result, the maximum number of images does not exceed 
thirteen for a square model. 

\begin{figure}[t]
\includegraphics[width=9.0cm]{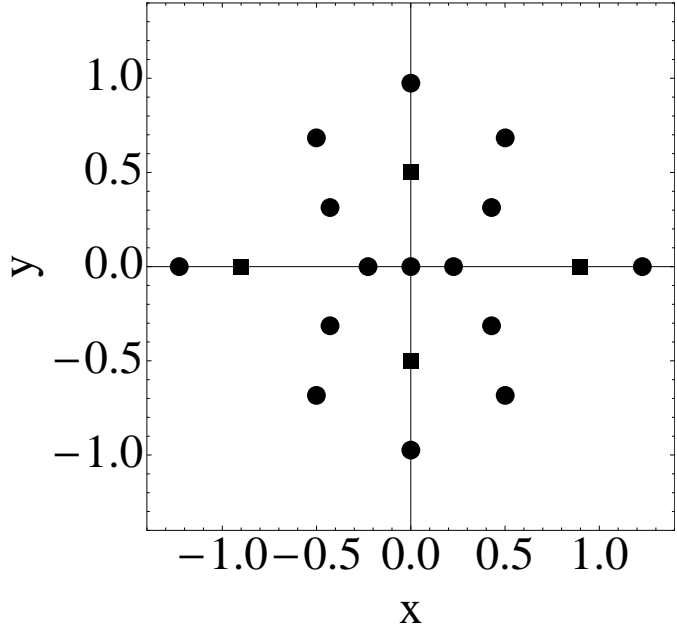}
\caption{
Four point mass lenses at the vertices of a rhombus 
and fifteen images: 
The source is located at the center. 
The filled squares and disks denote 
the positions of the lenses and images, respectively. 
We assume equal masses, and $k=0.9$ and $\ell=0.5$. 
}
\label{fig1}
\end{figure}

Figure $\ref{fig1}$ shows 
fifteen image positions due to four point masses 
at the vertices of a rhombus. 

\subsection{Five lenses at the center and vertices of a rhombus}
Let five masses located at the center and vertices of a rhombus, 
where the location of each mass  
is denoted as 
$\epsilon_1=k$, $\epsilon_2=i \ell$, 
$\epsilon_3=-k$, $\epsilon_4=-i \ell$, $\epsilon_5=0$ 
for $k, \ell \in R$. 
Without loss of generality, we assume $k\geq\ell$ again. 
By respecting symmetries regarding the principal axes 
of the rhombus, we take mass ratios as 
$m_1=m_3=\mu/2$, $m_2=m_4=\nu/2$ and $m_5=1-\mu-\nu$ 
for $0 < \mu < 1$, $0 < \nu < 1$ and $\mu + \nu < 1$. 

Let the source located at the center such that 
the equation can be rewritten as 
\begin{equation}
z z^* (z^{* 2}-k^2) (z^{* 2}+\ell^2) 
= 
z^{* 4} + [(1-\nu)\ell^2 - (1-\mu)k^2)]z^{* 2} 
- (1-\mu-\nu) k^2\ell^2 . 
\label{lenseq-five}
\end{equation}
We employ the polar coordinates as $z= r \exp(i\phi)$. 
Importantly, 
$r=0$ is no more root for Eq. ($\ref{lenseq-five}$), 
since this position is occupied by a lens mass.  
Henceforth, we shall seek roots for $r \neq 0$. 
Eq. ($\ref{lenseq-five}$) is reexpressed as 
\begin{equation}
r^6 - r^4 = A_5 e^{4i\phi} + B_5 e^{2i\phi} , 
\label{lenseq-five-r}
\end{equation}
where real quantities are defined as 
\begin{eqnarray}
A_5&=&k^2 \ell^2 [r^2 - (1-\mu-\nu)] , 
\label{A5}
\\
B_5&=& \left[ (1-\nu)\ell^2 - (1-\mu)k^2 \right] r^2 
- (\ell^2 - k^2) r^4 .  
\label{B5}
\end{eqnarray}
This has a structure very similar to Eq. ($\ref{lenseq-four-r}$). 
The L. H. S. of Eq. ($\ref{lenseq-five-r}$) is real and 
hence the R. H. S. must be real. 
We thus have three cases as 
$\exp(2i\phi)=\pm 1$ 
or $A_5[\exp(2i\phi)+\exp(-2i\phi)]+B_5=0$, 
which are investigated separately below. 

\noindent
\underline{Case 1.} $\exp(2i\phi)= 1$ ($\phi = 0$ or $\pi$):  \\ 
The lens equation becomes 
\begin{equation}
r^6 + (\ell^2 - k^2 -1) r^4 
- \left[ (1-\nu)\ell^2 - (1-\mu)k^2 + k^2 \ell^2 \right] r^2 
+ (1-\mu-\nu) k^2 \ell^2 = 0 , 
\label{lenseq-four-r1}
\end{equation}
where the coefficient of $r^4$ is always negative 
because of $k \geq \ell$ 
and the last constant term is positive 
for $\mu+\nu<1$. 
Roots for this equation cubic in $r^2$ are given 
by Cardano's method or alternatives \cite{Waerden}. 
By Descartes' rule of signs, 
we have the maximum number of positive roots 
as $2$, 
regardless of the sign of the coefficient of $r^2$. 

\noindent
\underline{Case 2.} $\exp(2i\phi)= -1$ ($\phi = \pi/2$ or $3\pi/2$): \\ 
The lens equation becomes 
\begin{equation}
r^6 - (\ell^2 - k^2 +1) r^4 
+ \left[ (1-\nu)\ell^2 - (1-\mu)k^2 - k^2 \ell^2 \right] r^2 
+ (1-\mu-\nu) k^2 \ell^2 = 0 , 
\label{lenseq-four-r2}
\end{equation}
where the last constant term is positive 
for $\mu+\nu<1$. 
This equation cubic in $r^2$ is solved  
by Cardano's method \cite{Waerden}. 
By Descartes' rule of signs, 
we have no positive root if both the coefficients 
of $r^4$ and $r^2$ are positive. 
Otherwise, it is possible that the equation admits  
two positive roots.  

\noindent 
\underline{Case 3}. $A_5[\exp(2i\phi)+\exp(-2i\phi)]+B_5=0$: \\ 
Given $r$, this case provides four roots of $\phi$ (mod $2\pi$). 
Eq. $(\ref{lenseq-five-r})$ is reduced to 
\begin{equation}
r^6 - r^4 + k^2 \ell^2 r^2 - (1-\mu-\nu) k^2 \ell^2 = 0 . 
\label{lenseq-four-r3}
\end{equation}
By Descartes' rule of signs, 
we have either three or one positive root. 

As a result, the number of images can reach  
$2\times 2 + 2\times 2 + 4\times 3 = 20$ 
if the conditions mentioned above are satisfied. 
The number twenty is exactly the maximum number of images for $N=5$. 

\begin{figure}[t]
\includegraphics[width=9.0cm]{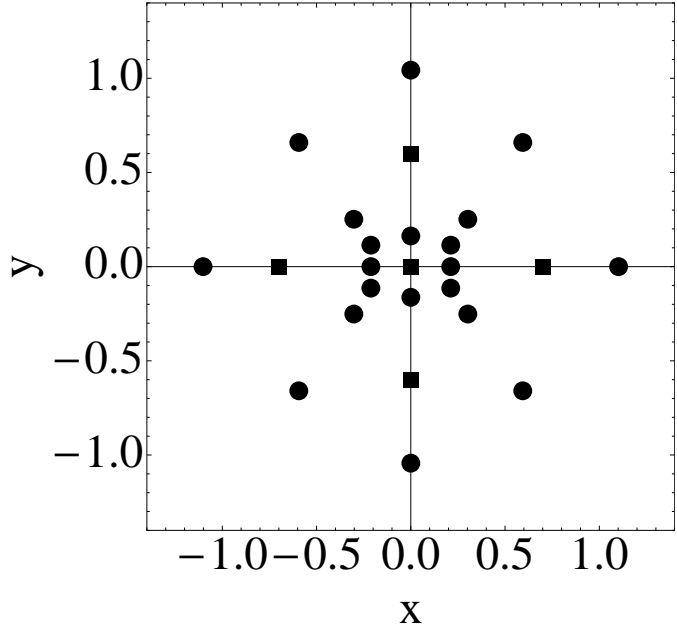}
\caption{
Five point mass lenses (filled squares) 
at the center and vertices of a rhombus 
and twenty images (filled disks): 
The source is located at the center. 
Here, $m_1 = m_2 = m_3 = m_4 = 6/25$, $m_5=1/25$, 
and $k=0.7$ and $\ell=0.6$. 
}
\label{fig2}
\end{figure}

Figure $\ref{fig2}$ shows 
twenty image positions due to five point masses 
at the center and vertices of a rhombus. 

\subsection{Approximate roots for small $|w| \neq 0$} 
Here, we comment on $w \neq 0$ cases. 
For arbitrary $w$,  the above models are not to be solvable by hand. 
For small $w$, however, one can construct approximate roots 
as follows.  
For the general form of the lens equation as 
$w=f(z, z^*)$, 
they are expressed in a form as 
\begin{equation}
z = z_0 + a w + b w^* +O(|w|^2) , 
\label{z}
\end{equation}
where $z_0$ denotes a root for $w=0$ (discussed above), and 
$a$ and $b$ are defined as 
\begin{eqnarray} 
a &=& 
\frac{1}{1-\left|\frac{\partial f}{\partial z^*}(z_0, z_0^*)\right|^2} , 
\label{a}
\\
b &=& 
-a^* \frac{\partial f}{\partial z^*}(z_0, z_0^*) . 
\label{b}
\end{eqnarray}
For our specific case as 
\begin{equation}
f(z, z^*) = z - \sum_{p=1}^N \frac{m_p}{z^* - \epsilon_p^*} , 
\label{f}
\end{equation}
we obtain 
\begin{eqnarray} 
a &=& 
\frac{1}{1-\left|\sum_p \frac{m_p}{(z_0^* - \epsilon_p^*)^2}\right|^2} , 
\label{a2}
\\
b &=& 
-\frac{\sum_p \frac{m_p}{(z_0^* - \epsilon_p^*)^2}}
{1-\left|\sum_p \frac{m_p}{(z_0^* - \epsilon_p^*)^2}\right|^2}. 
\label{b2}
\end{eqnarray}

It should be noted that these expressions are not always 
valid in the entire complex plane for $w$, 
because we assume a smooth one-to-one map between $w=0$ 
and $w \neq 0$ cases. This correspondence is singular when $1/a=0$. 
This is consistent with a fact that Eq. ($\ref{a2}$) 
is a magnification factor for an image at $z_0$ 
(i.e. the inverse of the Jacobian of the lens mapping). 
Figures 1 and 2 show non-merging images. 
Hence, the factor $a$ seems regular at $w=0$. 
Therefore, properties of the images for small $|w|$ 
may be similar to those for $w=0$. 
Further investigations of large $|w|$ are necessary for 
deeply understanding of the present lens model, 
especially image properties such as image shapes 
near critical curves.  
Finally, let us mention a single point-mass limit 
as $k, \ell \to 0$, for which 
Eqs. ($\ref{lenseq-four-r}$) and ($\ref{lenseq-five-r}$) 
reproduce the Einstein ring ($r=1$).

\section{Conclusion}
As an extension of four point mass lenses at the vertices 
of a rhombus, we have presented five point mass lenses 
at the center and vertices of a diamond, 
which constitute, for a source located behind the center, 
a soluble model giving expressions of all the image positions 
(with the maximum number of images as twenty for five lenses). 
For a source near the center,  
all the image positions have been obtained 
in the linear approximation. 
It is straightforward to extend this result 
to higher order approximations. 

The present model, though it is not realistic in astronomy,  
seems useful for theoretical investigations of lensing 
mostly because all the image positions (and thus magnification
factors) are given in terms of elementary functions. 
Comparing with a binary model $N=2$, 
a rhombus model with $N=4$ or $5$ 
has an advantage in the sense that 
one can control internal shear effects by choosing 
a ratio between two parameters $k$ and $\ell$ 
in addition to mass ratios. 
Further investigations are a topic of future work. 

Regular polygons cannot realize $5(N-1)$ images. On the other hand, 
four lenses on a rhombus and five ones in a rhombus can realize them. 
It is thus suggested that shear effects are important for the increase  
in image numbers. A squashed polygon may be an interesting 
configuration for lensing investigations.

\section*{Acknowledgements}
The author would like to thank the referee 
for useful comments.

%


\begin{thebibliography}{99}
  
\bibitem{WCW}
D. Walsh, R.~F. Carswell and R.~J. Weymann, 
Nature, {\bf 279} (1979), 381. 
\bibitem{FFKT}
M. Fukugita, T. Futamase, M. Kasai and E.~L. Turner, 
Astrophys. J. {\bf 393} (1992), 3. 
\bibitem{YKN}
C. Yoo, T. Kai, K. Nakao, Prog. Theor. Phys., 
{\bf 120} (2008), 937. 
\bibitem{MACHO} 
C. Alcock, et al., Nature, {\bf 365} (1993), 621. 
\bibitem{Asada09}H. Asada, Mon. Not. R. Astron. S. 
{\bf 394} (2009), 818. 
\bibitem{Asada02}H. Asada, Astron. Astrophys. {\bf 390} (2002), L11. 
\bibitem{Witt90}H. J. Witt, Astron. Astrophys. {\bf 236} (1990), 311. 
\bibitem{ES}H. Erdl and P. Schneider, 
Astron. Astrophys. {\bf 268} (1993), 453. 
\bibitem{SW}P. Schneider, and A. Weiss,  
Astron. Astrophys. {\bf 164} (1986), 237. 
\bibitem{Rhie01}S.~H. Rhie, (2001), arXiv:astro-ph/0103463. 
\bibitem{Rhie03}S.~H. Rhie, (2003), arXiv:astro-ph/0305166 
\bibitem{KN06}
D. Khavinson and G. Neumann,  
Proc. Amer. Math. Soc. {\bf 134}, (2006), 1077. 
\bibitem{MPW}S. Mao, A. Petters, H. J. Witt, (1997),  
 in {\it Proceedings of the Eighth Marcel Grossmann Meeting 
on General Relativity}, Ed. R. Ruffini, 
(Singapore, World Scientific) 
astro-ph/9708111. 
\bibitem{BKN}R. R. Bourassa, R. Kantowski, T. D. Norton, 
Astrophys. J. {\bf 185} (1973), 747. 
\bibitem{BK}R. R. Bourassa, R. Kantowski,  
Astrophys. J. {\bf 195} (1975), 13. 
\bibitem{Waerden}B. L. van der Waerden, 
{\it Algebra I} (Springer, 1966) 


\end{thebibliography}
\end{document}